\documentclass[prl,twocolumn,showpacs]{revtex4}

\newcommand{\boldvec}[1]{\mbox{\boldmath$#1$}}
\newcommand{\smallvec}[1]{\mbox{\boldmath$\scriptstyle#1$}}
\newcommand{\annhilate}[2]{\hat{#1}^{\phantom\dag}_{{#2}}}
\newcommand{\create}[2]{\hat{#1}^{\dag}_{{#2}}}

\usepackage{graphicx}

\begin{document}

\title{The role of boson-fermion correlations in the resonance theory
of superfluids}

\author{M. J. Holland}

\altaffiliation[Permanent address: ]{JILA, 440 UCB, University of
Colorado, Boulder, CO 80309-0440, U.S.A.}

\author{C. Menotti}

\author{L. Viverit} 

\altaffiliation[Alternative address: ]{Dipartimento di Fisica,
Universit\`a di Milano, via Celoria 16, I-20133 Milano, Italy.}

\affiliation{CRS-BEC INFM and Dipartimento di Fisica, Universit\`a di
Trento, via Sommarive 14, I-38050 Povo, Italy}

\date{\today}

\begin{abstract}

Correlations between a composite boson and a fermion pair are
considered in the context of the crossover theory of fermionic to
bosonic superfluidity. It is shown that such correlations are the
minimal ingredients needed in a many-body theory to generate the right
boson-boson scattering length in the Bose-Einstein limit of the
crossover.

\end{abstract}


\maketitle

In the quantum theory of many-body systems, superfluidity is connected
with a broken symmetry which is not anticipated from the underlying
microscopic Hamiltonian. There are two distinct types of
superfluidity. A first type occurs in fermionic systems in the
presence of an effectively attractive fermion-fermion interaction (as
in superconducting metals). The symmetry breaking there is associated
with the development of mean-field pair-correlations with coherence
length much larger than the interparticle spacing. A second type of
superfluidity occurs in bosonic systems (such as atomic condensates)
where the symmetry breaking is associated with the condensation of
preformed bosons with spatial size much smaller than the interparticle
spacing.  For many years it has been conjectured that these two types
of superfluidity are in fact alternative limits of a single universal
phenomenon, and that one could continuously pass from fermionic to
bosonic superfluidity by properly varying the fermion interaction
parameters.

This conjecture has attracted a great deal of attention. It has been
shown that the Bardeen-Cooper-Schrieffer (BCS) theory~\cite{bcs}
reduces to Bose-Einstein condensation (BEC) of bosonic dimers as the
pair correlation length becomes small compared with the interparticle
spacing~\cite{Leggett}. For a dilute Fermi gas, in which the mean
interparticle spacing is large compared to the extension of the
two-body potential, the simplest way to obtain the crossover is via a
scattering resonance. This corresponds to bringing the highest-lying
bound state from just above to just below threshold. In a dilute gas
the atom-atom interaction can be parametrized by the only nonvanishing
contribution to the two-body $T$-matrix at low energy, namely the
fermionic $s$-wave scattering length~$a_F$ (with $T=4\pi\hbar^2a_F/m$
and $m$ is the atomic mass). The crossover from fermionic to bosonic
superfluidity thus appears through the tuning of $a_F$ from negative
values, through infinity, to positive values. A series of improvements
to the basic theory to better account for the crossover region have
been developed and extensively studied during the last
decades~\cite{haussmann}.

It has recently been pointed out~\cite{strinati,petrov} that although
these theories model the passage from fermionic to bosonic
superfluidity, they have one major flaw in that they fail to reproduce
the correct equation of state of the system in the bosonic limit. This
is due to the fact that they predict the wrong value for the
boson-boson scattering length, $a_B=2a_F$. A solution of the
four-fermion Schr\"odinger equation for contact scattering in vacuum,
which in principle can be determined exactly, has recently been
found~\cite{petrov} and provided the value $a_B\simeq 0.6a_F$.

Progress in experiments with quantum atomic gases has lead to
experimentally accessible systems which can directly address the
crossover region using Feshbach scattering
resonances~\cite{expts}. Bose-Einstein condensation of two-fermion
dimers has been observed and the experimental study of the crossover
is actively being pursued.  The first data collected on the
Bose-Einstein condensed cloud were consistent with the
$0.6a_F$~\cite{grimm}. A consequence is that crossover theories that
predict in the BEC limit the value $2a_F$ for the boson-boson
scattering length will not yield the correct results for all
observables dependent on interactions, e.g.\ the collective modes, the
vortex core structure, the internal energy, and even the macroscopic
spatial extent of a confined cloud.

The aim of this paper is to present a theory of the crossover which
correctly reproduces the Gross-Pitaevskii and Bogoliubov theory of the
non-ideal Bose gas with the right boson-boson interaction. As we shall
see this theory will recover on the fermionic side the BCS theory
including the Gorkov corrections due to density and spin
fluctuations~\cite{gorkov}. We begin by pointing out the essential
ingredients which are missing in standard crossover theories and which
have to be included to achieve this goal. Our approach was motivated
by Ref.~\cite{petrov} where the few-body problem of four fermions
interacting in vacuum was considered. The essential property leading
to the correct boson-boson $T$-matrix was identified within the
contact scattering formalism as the three-particle correlation between
a composite boson and a fermion pair. This followed from the
observation that in the limit in which the distance ${\boldvec r_1}$
between any two of the fermions becomes small, the four-fermion
wavefunction must factorize as
\begin{eqnarray}
\Psi(\boldvec r_1,\boldvec r_2,\boldvec R)\rightarrow f(\boldvec
r_2,\boldvec R)(1/4\pi r_1-1/4\pi a_F),
\label{eqpetrov}
\end{eqnarray}
where ${\boldvec r_2}$ is the distance between the two other fermions
and ${\boldvec R}/\sqrt{2}$ the distance between the centers of mass
of the pairs. The function $f(\boldvec r_2,\boldvec R)$---an effective
three-body wavefunction of a composite boson and two
fermions---contains all information needed to determine the
boson-boson scattering length.  The reason why standard crossover
theories fail to give the correct $T$-matrix in the BEC limit is then
revealed, because they are typically based on the BCS assumption that
only two-fermion correlations are important, and only those are
withheld in the correlation hierarchy.

In this paper we show that a self-consistent many-body theory of the
crossover can be constructed which includes three-particle
correlations, and correctly reduces to the known results in the
appropriate limits. This modification not only brings about
quantitative corrections, but also introduces a qualitative change in
the crossover picture. A major consequence is that the superfluid
order parameter in this crossover theory is not given by the simple
integration of a BCS-type pairing field.

A system of fermions in two internal states ($\uparrow$ and
$\downarrow$) close to a Feshbach resonance can be described using the
many-body Hamiltonian~\cite{holland1,timmermans}:
\begin{eqnarray}
\label{Hambf}
{\hat H_{\rm res}}&=&\sum_{\smallvec
k,\sigma=\uparrow,\downarrow}\epsilon_{\smallvec k}
\create{a}{\smallvec k\sigma}\annhilate{a}{\smallvec k\sigma}
+\sum_{\smallvec q}\Bigl( \frac{\epsilon_{\smallvec q}}2+\nu\Bigr)
\create{b}{\smallvec q} \annhilate{b}{\smallvec q} \nonumber\\&&
+\sum_{\smallvec q\smallvec k} g_{\smallvec
k}\Bigl(\create{b}{\smallvec q} \annhilate{a}{\smallvec q/2-\smallvec
k\downarrow} \annhilate{a}{\smallvec q/2+\smallvec
k\uparrow}+\mbox{H.c.}\Bigr)
\end{eqnarray}
where $\epsilon_{\smallvec k}=\hbar^2k^2/2m$ is the free fermion
dispersion relation, $g_{\smallvec k}$ is the matrix element relating
two free fermions in the open channel to the closed channel bound
state near threshold, and $\nu$ is the bare detuning of the bound
state. The operators $\hat{a}_{\smallvec k\sigma}^{(\dagger)}$
annihilate (create) open channel fermions with momentum ${\bf k}$ and
spin $\sigma$, while $\hat{b}_{\smallvec k}^{(\dagger)}$ annihilate
(create) closed channel bosons.  We ignore here the direct interaction
of the fermionic degrees of freedom which would otherwise give rise to
an asymptotic background value for the scattering length, since that
plays no role at detunings sufficiently close to the resonance
value. This can nevertheless be reintroduced in a straightforward
manner as has been previously discussed~\cite{servaas}.

In resonance superfluidity theory the information contained in the
wavefunction $f({\bf r}_2,{\bf R})$ for a boson and a fermion pair is
in principle conveyed by the correlation function
\begin{equation}
\label{corr}
\bigl<
\annhilate{b}{-\smallvec q} \annhilate{a}{\smallvec q/2-\smallvec k
\downarrow}\annhilate{a}{\smallvec q/2+\smallvec k \uparrow}\bigr>.
\end{equation}
Notice, however, that the two quantities do not simply coincide
because the one in Eq.~(\ref{corr}) describes the correlations between
bare fermionic and bosonic states, while $f({\bf r}_2,{\bf R})$ gives
the correlations between states dressed by the interactions (and thus
physically observable). The connection between the two quantities is
not obvious and must be stated explicitly.

The relation can be well understood by considering the two-fermion
correlations with zero center of mass momentum, which in a vacuum
coincides with the Fourier transform of the relative wavefunction.
The bare correlation function is simply $\bigl<
\annhilate{a}{-\smallvec k\downarrow} \annhilate{a}{\smallvec
k\uparrow} \bigr>$. This can however never be an eigenstate of
Eq.~(\ref{Hambf}). An eigenstate can be constructed by considering the
following linear combination of $\bigl< \annhilate{a}{-\smallvec
k\downarrow} \annhilate{a}{\smallvec k\uparrow} \bigr>$ and
$\bigl<\annhilate{b}{0}{}\bigr>$ (dressed state):
\begin{equation}
\Psi_{\smallvec k}= \bigl< \annhilate{a}{-\smallvec k\downarrow}
\annhilate{a}{\smallvec k\uparrow} \bigr>+{\cal P} \frac{g_{\smallvec
k}}{2\epsilon_{\smallvec k}-E} \bigl<\annhilate{b}{0}{}\bigr>
\label{ansatztwo}
\end{equation}
with $\cal P$ denoting the Cauchy Principal Value and $E$ a solution
of
\begin{equation}
\label{Esum}
E=\nu-{\cal P}\sum_{\smallvec k}\frac{g_{\smallvec
k}^2}{2\epsilon_{\smallvec k}-E}\,.
\end{equation}
The nature of the solution depends on the presence or absence of a
bound state indicated by the sign of the renormalized detuning
$\bar\nu$. This is defined as
\begin{equation}
\bar\nu=\nu-\sum_{\smallvec k}\frac{g_{\smallvec
k}^2}{2\epsilon_{\smallvec k}}
\label{renorm}
\end{equation}
and is physically related to the magnetic field shift from the
Feshbach resonance~\cite{servaas}. The case of $\bar\nu<0$ corresponds
to the BEC side of the resonance. There a bosonic dimer bound state
exists and the solution of Eq.~(\ref{Esum}) coincides with the bound
state energy $E=-\hbar^2/ma_F^2$~\cite{holland,stoof,pethick}. For
$\bar\nu>0$ there is no bound state and the solution is
$E=\bar\nu$. Evolving Eq.~(\ref{ansatztwo}) for two particles in a
vacuum under the Hamiltonian Eq.~(\ref{Hambf}) gives the Schr\"odinger
equation
\begin{equation}
i\hbar\frac{d\Psi_{\smallvec k}}{dt}=2\epsilon_{\smallvec k}
\Psi_{\smallvec k}+\sum_{\smallvec k'}U_{\smallvec k,\smallvec k'}
\Psi_{\smallvec k'}
\label{schro}
\end{equation}
with real separable potential
\begin{equation}
U_{\smallvec k,\smallvec k'}={\cal P}\frac{g_{\smallvec
k}g_{{\smallvec k'}}}{2\epsilon_{\smallvec k}-E}.
\label{veff}
\end{equation}
What remains is to show that this potential generates the correct
scattering length at all detunings~$\bar\nu$. To this end we must
obtain the $T$-matrix by solving the Lippmann-Schwinger equation as
derived from Eq.~(\ref{schro})~\cite{pethicksmith}
\begin{equation}
T_{\smallvec k,\smallvec k'}= U_{\smallvec k,\smallvec k'}
+\sum_{\smallvec q}\frac{U_{\smallvec k,\smallvec q}T_{\smallvec
q,\smallvec k'} }{2\epsilon_{\smallvec k'}-2\epsilon_{\smallvec
q}+i\delta},\quad \delta\rightarrow0^+
\label{lipp}
\end{equation}
in the limit of zero scattering energy. Substituting Eq.~(\ref{veff})
into Eq.~(\ref{lipp}) and performing the integration, we recover the
desired result for the zero-energy $T$-matrix
\begin{equation}
T=-\frac{g_0^2}{\bar\nu}.
\label{tform}
\end{equation}
Eq~(\ref{tform}) provides the correct behavior of the tuning of the
scattering length around resonance~\cite{servaas}, with the usual
definition $T=4\pi\hbar^2a_F/m$, and confirms that the potential~in
Eq.~(\ref{veff}) leads to the correct effective fermion interaction
properties.

The key point is that Eq.~(\ref{schro}) is also the time-dependent
Schr\"odinger equation for the relative wavefunction
$\bigl<\annhilate{\alpha}{-\smallvec
k\downarrow}\annhilate{\alpha}{\smallvec k\uparrow}\bigr>$ of two
particles in a vacuum evolving under the single-channel Hamiltonian
\begin{eqnarray}
&&{\hat H}_{\rm single}=\sum_{\smallvec k\sigma}\epsilon_{\smallvec k}
\create{\alpha}{\smallvec k\sigma}\annhilate{\alpha} {\smallvec
k\sigma}\nonumber\\ &+&\sum_{\smallvec q\smallvec k\smallvec k'}
U_{\smallvec k,\smallvec k'}\, \create{\alpha}{\smallvec q/2+\smallvec
k\uparrow} \create{\alpha}{\smallvec q/2-\smallvec k\downarrow}
\annhilate{\alpha}{\smallvec q/2-\smallvec k'\downarrow}
\annhilate{\alpha}{\smallvec q/2+\smallvec k'\uparrow}
\label{Heff}
\end{eqnarray}
Introducing the dressed wavefunction Eq.~(\ref{ansatztwo}) is
therefore equivalent to eliminating the bosonic degree of freedom from
the theory by introducing effective purely fermionic quantities. We
wish to point out that there is an alternative method of eliminating
the bosonic degrees of freedom based on the application of the
Hubbard-Stratonovich transformation to the functional integral
expression for the partition function~\cite{popov}. This is distinct
from our approach, because it leads to a formal result in which the
effective fermion-fermion interaction potential contains bare
quantities, namely $\nu$. In contrast the potential we give here is a
real effective potential which contains only renormalized quantities,
$E$ or $\bar\nu$.

The implication of the above considerations goes beyond what we have
presented so far, because the dressing procedure can be extended to
treat correlation functions of any number of particles. It is now
possible to make the direct link between the bare and the dressed
correlation functions for a composite boson and a fermion pair,
Eq.~(\ref{corr}) and $f(\boldvec r_2,\boldvec R)$. This can only be
done assuming the contact scattering form for the boson-fermion
vertex, i.e. that $g_{\smallvec k}=g$ independent of $\boldvec
k$. Since this assumption leads to an ultraviolet divergent theory, we
need to introduce a momentum cutoff $K$ to the wavevector
sums~\cite{servaas}. This implies for instance that the renormalized
detuning in Eq.~(\ref{renorm}) is given by $\bar\nu=\nu-mK
g^2/2\pi^2\hbar^2$. Recursively applying the same procedure used for
the dressing of the two-fermion wavefunction, we are led to construct 
an ansatz for the four-fermion wavefunction:
\begin{eqnarray}
&&
\Psi_{\smallvec{k_1},\smallvec{k_2}, \smallvec{q}}=
\langle
\annhilate{a}{- \smallvec q/2-\smallvec k_1 \downarrow}
\annhilate{a}{- \smallvec q/2+\smallvec k_1 \uparrow}
\annhilate{a}{\smallvec q/2-\smallvec k_2 \downarrow}
\annhilate{a}{\smallvec q/2+\smallvec k_2 \uparrow}
\rangle
\nonumber\\
&&\quad
+\beta_{\smallvec k_1}
\langle
\annhilate{b} {- \smallvec q}
\annhilate{a}{\smallvec q/2-\smallvec k_2 \downarrow}
\annhilate{a}{\smallvec q/2+\smallvec k_2 \uparrow}
\rangle \nonumber\\
&&\quad
+\beta_{\smallvec k_2}
\langle
\annhilate{b} {\smallvec q}
\annhilate{a}{- \smallvec q/2-\smallvec k_1 \downarrow}
\annhilate{a}{- \smallvec q/2+\smallvec k_1 \uparrow}
\rangle \nonumber\\
&&\quad
-\beta_{(\smallvec q-\smallvec k_1-\smallvec k_2)/2}
\langle
\annhilate{b} {\smallvec k_1- \smallvec k_2}
\annhilate{a}{-\smallvec q/2-\smallvec k_1 \downarrow}
\annhilate{a}{\smallvec q/2+\smallvec k_2 \uparrow}
\rangle\nonumber\\
&&\quad
-\beta_{(\smallvec q+\smallvec k_1+\smallvec k_2)/2}
\langle
\annhilate{b} {\smallvec k_2- \smallvec k_1}
\annhilate{a}{\smallvec q/2-\smallvec k_2 \downarrow}
\annhilate{a}{\smallvec -q/2+\smallvec k_1 \uparrow}
\rangle\nonumber\\
&&\quad
+\beta_{\smallvec k_1}
\beta_{\smallvec k_2}
\langle
\annhilate{b}{- \smallvec q} \annhilate{b}{\smallvec q} \rangle
\nonumber\\
&&\quad
-\beta_{(\smallvec q-\smallvec k_1-\smallvec k_2)/2}
\beta_{(\smallvec q+\smallvec k_1+\smallvec k_2)/2}
\langle
\annhilate{b} {\smallvec k_2 - \smallvec k_1}
\annhilate{b} {\smallvec k_1- \smallvec k_2}
\rangle
\label{ansatzfour}
\end{eqnarray}
where $\beta_{\smallvec k} ={\cal P}\{g/(2\epsilon_{\smallvec
k}-E)\}$. It can be shown that for four-particles in vacuum the
evolution of this ansatz under the resonance Hamiltonian
Eq.~(\ref{Hambf}) is equivalent to the evolution of the four-particle
wavefunction
\begin{equation} 
\bigl< \annhilate{\alpha}{- \smallvec q/2-\smallvec k_1 \downarrow}
\annhilate{\alpha}{- \smallvec q/2+\smallvec k_1 \uparrow}
\annhilate{\alpha}{\smallvec q/2-\smallvec k_2 \downarrow}
\annhilate{\alpha}{\smallvec q/2+\smallvec k_2 \uparrow} \bigr>
\label{fourwf}
\end{equation}
under the single channel Hamiltonian Eq.~(\ref{Heff}), with the
potential $U_{\smallvec k,\smallvec k'}$ given in Eq.~(\ref{veff}).
Since we have proved that this interaction leads to the correct
two-body $T$-matrix, this result is crucial because it is exactly the
same equation which was solved in Ref.~\cite{petrov} to obtain the
correct value of the boson-boson scattering length. It should also be
clear that the procedure we have outlined can be systematically
extended to incorporate more and more particles.

We are now in a position to make a definitive statement about the
complete set of correlation functions in the resonance theory needed
to reconstruct the information represented by $f(\boldvec r_2,\boldvec
R)$. Having two fermions approaching each other is equivalent to
summing over one of the two relative momenta, e.g.\ $\boldvec k_1$, in
Eq.~(\ref{fourwf}) or equivalently in Eq.~(\ref{ansatzfour}). In this
way we formulate the complete set:
\begin{equation}
\begin{array}{l}
\langle \annhilate{b}{-\smallvec q}
\annhilate{b}{\smallvec q} \rangle\,,\\
\langle \annhilate{b}{-\smallvec q} \annhilate{a}{\smallvec
q/2-\smallvec k_2 \downarrow} \annhilate{a}{\smallvec q/2+\smallvec
k_2 \uparrow} \rangle\,,\\
\sum_{\smallvec
k_1} \langle \annhilate{a}{- \smallvec q/2-\smallvec k_1 \downarrow}
\annhilate{a}{- \smallvec q/2+\smallvec k_1 \uparrow}
\annhilate{a}{\smallvec q/2-\smallvec k_2 \downarrow}
\annhilate{a}{\smallvec q/2+\smallvec k_2 \uparrow}\bigr>.
\end{array}
\label{corrlist}
\end{equation}
The desired many-body theory must therefore include these correlation
functions in order to generate the correct equation of state in the
bosonic limit. We can now proceed to construct an approximate
many-body theory using the standard methods of cumulant
expansion~\cite{manybody} so that we keep explicitly the correlation
functions in Eq.~(\ref{corrlist}) and all the correlation functions of
the same order. The key point is that the resulting theory closes
within a set of functions which depend on a maximum of two vector
arguments (e.g.\ the functions given in Eq.~(\ref{corrlist}) depend at
most on $\boldvec q$ and $\boldvec k_2$). At a computational level,
this is an essential point, because a two vector field in a
translationally invariant system contains three nontrivial degrees of
freedom, and a three dimensional theory is tractable. This is the
minimal complexity because it is consistent with the intrinsic
dimensionality of $f(\boldvec r_2,\boldvec R)$ which encapsulates the
few-body scattering calculation that this theory must reduce to in the
limit of zero density. We also point out that this is an enormous
simplification over the result of a direct application of the cumulant
expansion to keep all correlation functions of order four fermions and
below, i.e.\ including those of the form given in
Eq.~(\ref{fourwf}). Such a function depends on three vectors, and even
in a uniform system, where rotational symmetries can be exploited,
contains at least six nontrivial degrees of freedom.

It must be emphasized that the possibility of reducing the number of
degrees of freedom in the whole crossover region is peculiar to the
resonance Hamiltonian. This is because in order to close the set of
equations in the many-body theory with three-particle correlation
functions, it is crucial to apply the contact scattering
approximation. In the resonance theory, Eq.~(\ref{Hambf}), this
implies $g_{\smallvec k}\rightarrow g$ independent of $\boldvec
k$. However, the resulting effective potential Eq.~(\ref{veff}) has a
residual $\boldvec k$-dependence which makes the potential and thus
the theory well-defined even at $\bar\nu=0$. However, in the single
channel formulation, Eq.~(\ref{Heff}), this point causes problems as
the scattering length $a_F\rightarrow\infty$ and the two-body
$T$-matrix diverges. For this region there is no obvious way of
incorporating correctly correlations of more than two particles in the
many-body theory in a pseudopotential approximation. One can of course
utilize a nonlocal potential and keep the full four-particle functions
of the form in Eq.~(\ref{fourwf}) in that case.

We now consider explicitly the consequence of the above formalism in
the bosonic and fermionic superfluidity limits. The complete theory of
the dilute nonideal Bose gas in the bosonic limit emerges naturally
since it depends only on the eigensolution of the two-body problem and
the interactions of the resulting dimers. Both of these elements we
have discussed in depth. Thus, when the usual approximations are made,
the Gross-Pitaevskii equation, the Bogoliubov theory, and the Popov
theory, will emerge. In the fermionic limit, the set of correlation
functions that we are keeping leads to the possibility of constructing
a pair wavefunction that correctly includes the full many-body
effects. In particular one can prove that the following ansatz:
\begin{eqnarray}
\label{pairansatz}
&& \hspace*{-2pc}\Psi_{\smallvec k}= \bigl< \annhilate{a}{-\smallvec
k\downarrow} \annhilate{a}{\smallvec k\uparrow} \bigr>
-\sum_{\smallvec q}\beta_{\smallvec q/2+\smallvec k}
\Bigl(\bigl<\annhilate{b} {\smallvec q} \create{a}{\smallvec
q+\smallvec k\uparrow} \annhilate{a}{\smallvec k\uparrow}\bigr>
\nonumber\\ &&\hspace*{-2pc} \qquad - \bigl<\annhilate{b}{-\smallvec q}
\annhilate{a}{-\smallvec k\downarrow}\create{a}{-\smallvec q-\smallvec
k\downarrow} \bigr>\Bigr),
\end{eqnarray}
evolves using Eq.~(\ref{Hambf}) as
$\bigl<\annhilate{\alpha}{-\smallvec
k\downarrow}\annhilate{\alpha}{\smallvec k\uparrow}\bigr>$ evolves
under the Hamiltonian in Eq.~(\ref{Heff}):
\begin{eqnarray}
&& \hspace*{-2pc} i\hbar\frac{d\bigl<\annhilate{\alpha}{-\smallvec
  k\downarrow}\annhilate{\alpha}{\smallvec k\uparrow}\bigr>}
  {dt}=2\epsilon_{\smallvec k}\bigl<\annhilate{\alpha}{-\smallvec
  k\downarrow}\annhilate{\alpha}{\smallvec k\uparrow}\bigr>\nonumber\\
  && \hspace*{-2pc}-\sum_{\smallvec q\smallvec k'}U_{\smallvec
  q/2+\smallvec k,\smallvec k'} \Bigl(\bigl<\create{\alpha}{\smallvec
  q+\smallvec k\uparrow}\annhilate{\alpha}{\smallvec k\uparrow}
  \annhilate{\alpha}{\smallvec q/2-\smallvec k'
  \downarrow}\annhilate{\alpha}{\smallvec q/2+\smallvec
  k'\uparrow}\bigr>\nonumber\\ && \hspace*{-2pc}\qquad -
  \bigl<\annhilate{\alpha}{-\smallvec k
  \downarrow}\create{\alpha}{-\smallvec q-\smallvec
  k\downarrow}\annhilate{\alpha}{-\smallvec q/2-\smallvec
  k'\downarrow}\annhilate{\alpha}{-\smallvec q/2+\smallvec
  k'\uparrow}\bigr>\Bigr).
\label{singlepair}
\end{eqnarray}
Eq.~(\ref{singlepair}) shows that Eq.~(\ref{pairansatz}) contains the
full pair wavefunction in the medium and not just the BCS approximate
factorized form. Notice that an analogous ansatz and correspondence
can be formulated for the density correlation function
$\bigl<\create{\alpha}{\smallvec
k\uparrow}\annhilate{\alpha}{\smallvec k \uparrow}\bigr>$. Together
these imply that we not only recover the complete BCS equations but
also the effects beyond mean-field such as the Gorkov corrections to
the superfluid gap.

In conclusion, a consistent theoretical framework has been presented
for the crossover of superfluidity from the fermionic to the bosonic
type. The inclusion of the correlation function between a boson and a
fermion pair was required to reproduce the correct bosonic equation of
state. This has lead us to develop a fundamental change to the picture
of the crossover physics.

We would like to thank S. Stringari, L. Pitaevskii, and S. Giorgini
for discussions. L.V. acknowledges support from Universit\`a di Milano
and C.M. and L.V. from CRS-BEC Trento. M.H. acknowledges support from
the National Science Foundation and from the U.S. Department of
Energy, Office of Basic Energy Sciences via the Chemical Sciences,
Geosciences and Biosciences Division.

\end{document}